# High crystalline quality single crystal CVD diamond

**P M Martineau[1], M P Gaukroger[1], K B Guy[1], S C Lawson[1], D J Twitchen[2], I Friel[2], J O Hansen[3], G C Summerton[3], T P G Addison[3] and R Burns[3]**

[1]DTC Research Centre, Belmont Road, Maidenhead SL6 6JW, UK
[2]Element Six, King's Ride Park, Ascot SL5 8BP UK
[3]Element Six, Johannesburg, SA

**Abstract.** Homoepitaxial chemical vapour deposition (CVD) on high pressure high temperature (HPHT) synthetic diamond substrates allows production of diamond material with controlled point defect content. In order to minimize the extended defect content, however, it is necessary to minimize the number of substrate extended defects that reach the initial growth surface and the nucleation of dislocations at the interface between the CVD layer and its substrate. X-ray topography has indicated that when type IIa HPHT synthetic substrates are used the density of dislocations nucleating at the interface can be less than 400 cm$^{-2}$. X-ray topography, photoluminescence imaging and birefringence microscopy of HPHT grown synthetic type IIa diamond clearly show that the extended defect content is growth sector dependent. <111> sectors contain the highest concentration of both stacking faults and dislocations but <100> sectors are relatively free of both. It has been shown that HPHT treatment of such material can significantly reduce the area of stacking faults and cause dislocations to move. This knowledge, coupled with an understanding of how growth sectors develop during HPHT synthesis, has been used to guide selection and processing of substrates suitable for CVD synthesis of material with high crystalline perfection and controlled point defect content.
**PACS** 61.72.Lk, 61.72.Ff, 61.72.Cc, 61.72.Jn

## 1. Introduction

Production of synthetic diamond material for optical and electronic applications requires the control of the point defects that influence the relevant properties of the material. Extended defects, such as dislocations, can also affect properties, such as birefringence, that are relevant in optical applications that include Raman lasers (Mildren et al 2008) and intra-cavity heat spreaders in solid state disc lasers (Millar et al 2008, Friel et al 2009). It has also been suggested that for use in some electronic devices, such as those designed to operate at high powers, diamond with very low dislocation density will be necessary (Umezawa et al 2008).

It is extremely difficult to control the impurity content of HPHT synthetic diamond to the extent necessary for all electronic and some optical applications. For CVD synthetic diamond material, however, such control has been demonstrated for both boron-doped (Scarsbrook et al 2002, Suzuki et al 2006) and phosphorus-doped (Koizumi and Suzuki 2006) diamond and also for material of high purity (Scarsbrook et al 2001, Isberg et al 2002, Teraji and Ito 2004, Tallaire et al 2005, Tranchant et al 2007, Achard et al 2007). Reduction of the dislocation content of such material has remained challenging however. Previous studies of dislocations in single crystal CVD diamond have shown that they have a tendency to nucleate at or near the interface with the substrate on which the material is homoepitaxially grown (Martineau et al 2004). X-ray topography has been used to characterize dislocations in layers grown on {001} substrates (Gaukroger et al 2008). They were found to be either edge or mixed 45 degree <001> dislocations, with sources for the latter type being associated with substrate polishing damage. The edge dislocations were found to nucleate in clusters at isolated points at the substrate interface, with such clusters having a strong effect on birefringence.

Extended defect content is of interest to gemmologists concerned with verifying whether diamond is natural or synthetic. Martineau et al. (2004) reported that dislocations grown into CVD single crystal diamond produced on {001} substrates tend to have line directions that are approximately parallel to the <001> growth direction and that this causes the material to show strain-related birefringence with a characteristic anisotropy, being most obvious for a viewing direction parallel to the growth direction. More recently it has been shown (Martineau et al 2009) that step flow growth can cause deviations of the line direction from <001>. Edge and mixed <001> dislocations have been modelled theoretically (Fujita et al 2006, Fujita et al 2007) and this work has been extended to modelling of the birefringence caused by clusters of edge dislocations (Pinto and Jones 2008). In contrast, type IIa natural diamond contains relatively high densities (up to $10^9$ - $10^{10}$ cm$^{-2}$) of dislocations (Paly'anov et al 1997, Willems et al 2006) resulting from plastic deformation and arranged either in slip bands or in geometric arrangements resulting from polygonization (Martineau et al 2004) that took place over the extremely long times during which diamonds resided in the earth's upper mantle.

In this paper we report the results of X-ray topography studies of both HPHT and CVD synthetic diamond that indicate how, by careful selection of type IIa HPHT substrate material combined with the CVD synthesis methods already developed, it is possible to grow diamond material that has not only controlled point defect content but also a low extended defect content (Twitchen et al 2007). As the ability to produce CVD single crystal diamond with controlled point defect content has already been demonstrated, this paper will focus on reduction of the concentration of extended defects in this material.

## 2. Experimental

### 2.1. HPHT-grown synthetic diamond substrates

The HPHT synthetic diamond substrates used and studied in this work were processed from type IIa samples produced using the temperature gradient method, starting from carefully selected as-grown synthetic type Ib HPHT diamond crystals that acted as seeds for growth of large single crystal type IIa samples with dimensions of several millimetres. The selection of the samples themselves, and also the regions within them suitable for use as substrates, was guided by X-ray topography, knowledge of the growth sector dependence of the extended defect content and knowledge of the way growth sectors developed during the synthesis process to give the observed sample morphologies, as illustrated in figure 1.

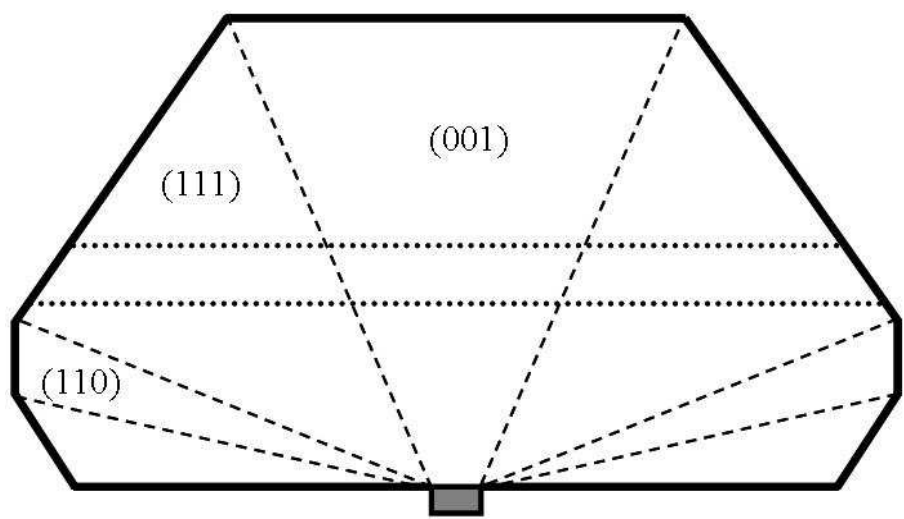

**Figure 1** A schematic diagram showing a {110} cross-sectional view of growth sectors in a type IIa HPHT synthetic diamond sample produced from a (001) seed. Broken lines represent growth sector boundaries and dotted lines indicate how a (001) plate might be cut.

Parallel-sided plates, subsequently used as substrates for CVD and for investigation of the effect of HPHT treatment, were processed by a combination of laser sawing and polishing. Some of the HPHT synthetic diamond was HPHT treated under diamond stabilizing pressure at a temperature between of 2400°C for 15 minutes in order to examine the effect on stacking faults and dislocations. Samples were re-polished after the HPHT treatment to remove any near-surface damage generated in during the treatment. All the single crystal CVD samples reported here were grown homoepitaxially on the {001} type IIa HPHT synthetic diamond substrates described above from plasmas generated using microwaves in a methane and hydrogen gas atmosphere.

UV-excited luminescence images were recorded using a DiamondView$^{TM}$. In this instrument samples were excited at wavelengths shorter than 228 nm using a filtered xenon arc lamp source. The use of above band gap radiation allows luminescence from the near-surface region to be selectively excited, giving images that clearly show up the way growth sectors with different impurity concentrations break the surface. More information about DiamondView$^{TM}$ can be found in Martineau et al. 2004.

X-ray topographs were recorded using a Lang camera fitted to a rotating anode X-ray generator with a molybdenum target, zirconium filter and 0.2 mm x 0.2 mm apparent source size. The operating conditions of 40 kV, 40 mA produced a bright source of molybdenum K$\alpha_1$ X-rays at 17.4 keV. A 250 µm slit collimator was used at a distance of 0.7 m from the source. Preliminary X-ray topographs were generated rapidly on dental film before high quality topographs, requiring longer exposure times, were recorded on nuclear plates (Ilford L4) and processed using standard techniques.

To provide information about the dislocations and stacking faults throughout a sample, projection topographs were recorded by translating the sample through the beam in order to expose its complete volume. The film was simultaneously translated to keep it in the same position relative to the sample. Exposure times were typically in the region of four hours but depended on sample size, the intensity of the reflection used and the degree of lattice perfection.

Various different reflections were used to generate projection topographs. For analysis of stacking fault contrast in HPHT synthetic diamond substrates, {111}, {220} and {422} reflections were used. Stacking faults in diamond tend to lie on {111} but, for a given {220} reflection, stacking faults will only be observed for two of the {111} planes because for the other two the fault vector lies in the diffraction plane. For the same reason, one set of stacking faults is absent for a {422} reflection. One set of stacking faults is absent for a {111} reflection, but for the different reason that the fault vector is parallel to the diffraction vector and has a magnitude that is an integral number of lattice spacing. {400} reflections were used to create topographs showing general views of the substrates showing stacking faults in all of the {111} planes with equal contrast.

Use of {111} diffraction had other advantages. For diamond, the {111} reflection gives the highest intensity and the exposure time was therefore minimized. The {111} reflections provide four equivalent views of the parallel-sided {001} plates studied in this research. With molybdenum K$\alpha_1$ X-rays, in each case the topograph presents a view in a direction that is inclined at 25.34° to the normal to the major faces of the sample. This angle is large enough to allow the entry and exit points of dislocations with line directions close to <001> to be clearly distinguished but it is small enough not to give rise to significant confusion from overlapping lines.

Quantitative birefringence microscopy was performed using a Metripol instrument based on a system developed by Glazer et al (1996). For parallel sided samples, the Metripol allows maps of |sin δ| to be created, where δ is the phase shift generated by the sample (of thickness

d) between components of the light passing through it with polarizations along the fast and slow axes, due to birefringence Δn, which can be determined from $\delta = (2\pi/\lambda)\Delta nd$. Monochromatic (λ = 550 nm) light was selected using a filter.

**3. Results**

The DiamondView luminescence images of a (001) parallel-sided type IIa HPHT synthetic diamond plate in figure 2 show a central (001) sector that is clearly delineated from the surrounding {111} sectors that show stronger luminescence.

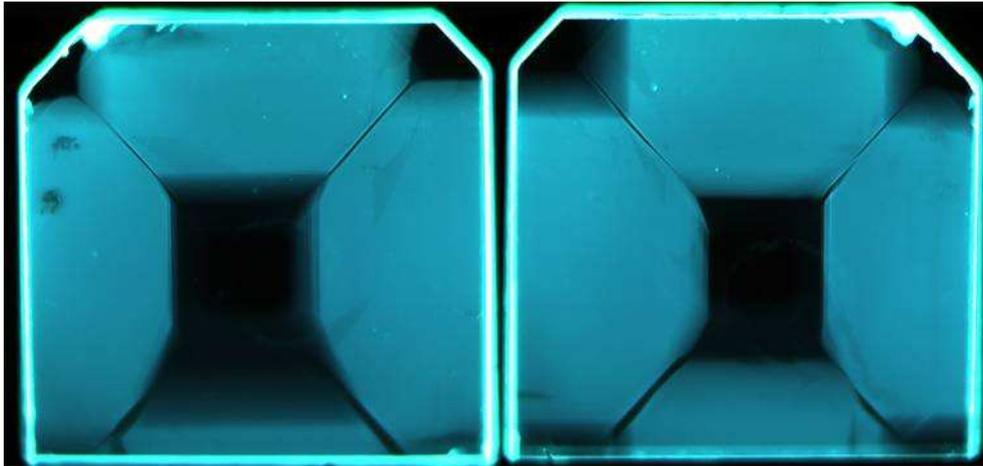

**Figure 2** DiamondView luminescence images of the two sides of a parallel-side type IIa HPHT synthetic diamond plate.

Figure 3 shows a (004) projection topograph of the same sample. Comparison of figures 2 and 3 indicates that stacking faults and dislocations are located predominantly in the {111} sectors, with the central (001) sector being almost entirely free of extended defects. Figure 1 illustrates growth sector development in relation to both the seed from which growth started and the final growth morphology. It also shows the approximate position from which the plate shown in figures 2 and 3 was cut to give close to the maximum possible area.

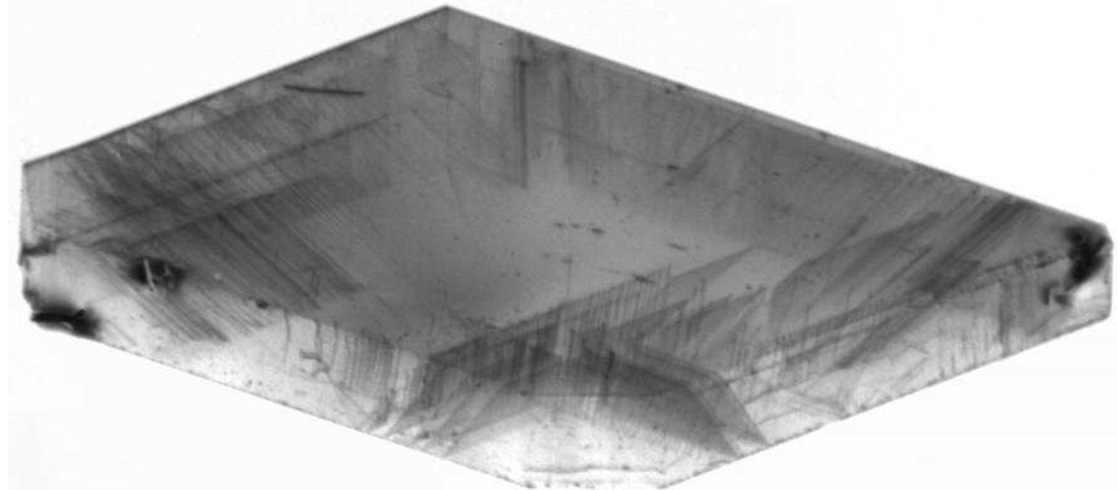

**Figure 3** (004) X-ray projection topograph of the same (001) HPHT synthetic diamond plate for which DiamondView images are shown in figure 2.

The topograph shown in figure 3 aids the visualisation of the distribution of the extended defects in the sample and from this topograph it can also be seen that each stacking fault lies in one of the four {111} planes. From their orientation it is clear that, in the sample from which the plate was produced, they radiated outwards from the direction of the seed towards the final {111} surfaces of the as-grown sample.

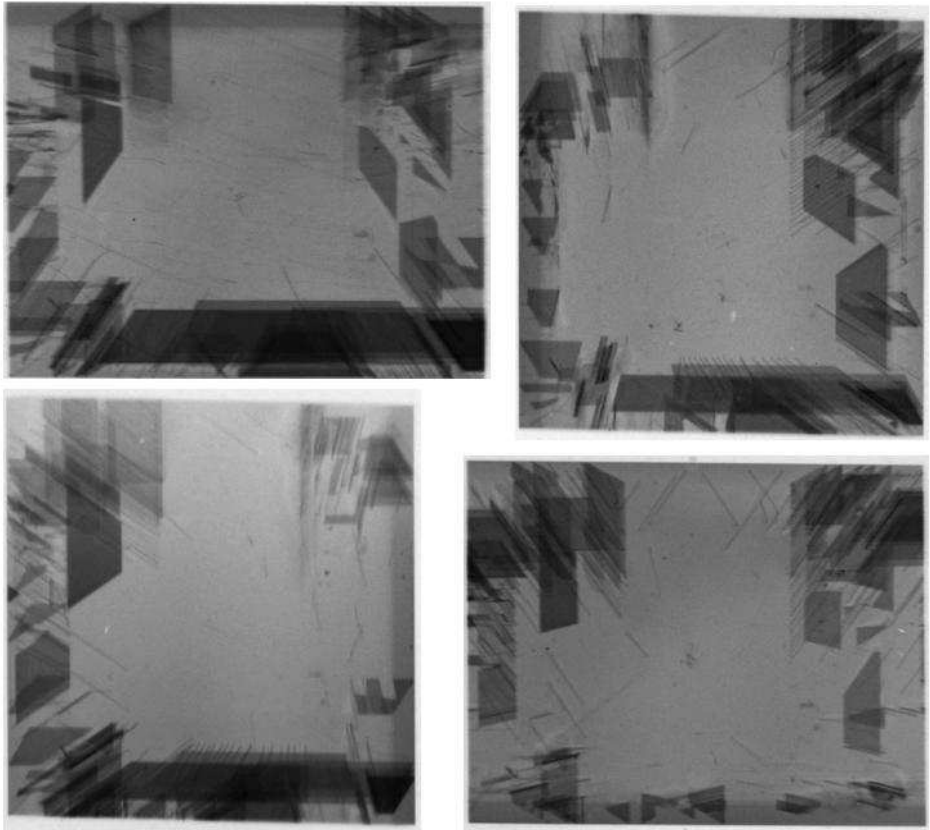

**Figure 4** {111} X-ray projection topographs of the HPHT synthetic diamond plate for each of the four <111> diffraction vectors.

When projection topographs were recorded for each of the four possible <111> diffraction vectors (figure 4), it was found that for each of the topographs one set of stacking faults was absent and that the corresponding fault plane was parallel to the diffraction plane. As mentioned above, stacking fault contrast is expected to be absent if the fault vector is parallel to the diffraction vector (<111> in this case) and equal to an integral number of lattice spacings. To check this interpretation, another set of four topographs of the same sample was recorded for four different <022> diffraction vectors, each making an angle of 45 degrees relative to [001]. In each of these topographs, contrast was present for two sets of stacking faults but absent for the other two. This was again consistent with a <111> fault vector perpendicular to the plane of the fault but, in this case, where contrast from faults was absent it was because the fault vector lay in the diffraction plane. For each {022}, there are two <111> directions that lie in the plane and two that lie out of it. As a final check, a further set of four topographs was recorded for four different <224> diffraction vectors and, as expected, in each topograph contrast was strong for one set of stacking faults, moderately strong for two sets and completely absent for the fourth set.

Figure 5 shows a set of four <111> projection topographs of a CVD layer on its type IIa HPHT synthetic substrate. The edges of the sample were processed after growth to remove regions of lower quality material so that contrast from defects within the sample could be seen more clearly. It was immediately apparent from these topographs that, where stacking faults broke the surface of the substrate, lines of dislocations had nucleated and propagated in the growth direction as the CVD process proceeded.

Figure 6 shows a pair of <111> projection topographs of the same sample that, because of the different projection, illustrate more clearly than figure 5 that the central region of the CVD layer is, like the corresponding region of the substrate, remarkably free of contrast relating to extended defects. The low dislocation density in this region of the samples is in marked contrast to that in the outer region of the sample and also in samples that we have studied that have been grown on type Ib HPHT synthetic substrates. It was clear that the substrate stacking faults had acted as sources of dislocations that reduced the crystal quality in the CVD layer that grew above them and a study of whether stacking faults could be annealed out was therefore carried out.

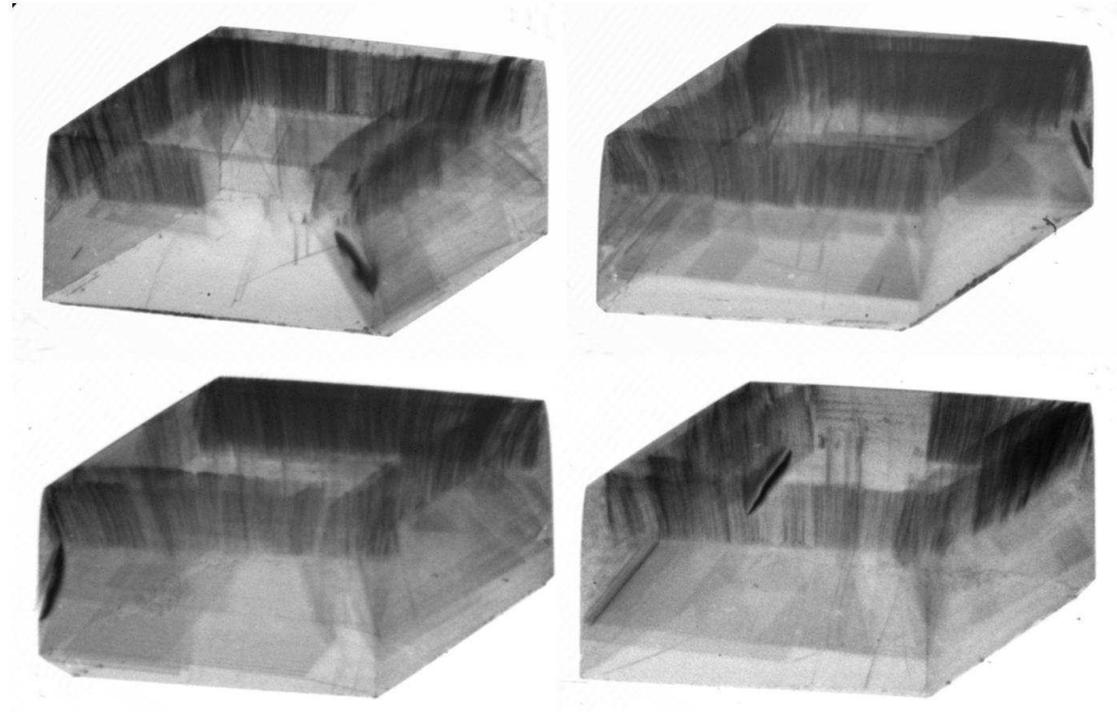

**Figure 5** Four {111} X-ray projection topographs of a CVD synthetic diamond layer on its type IIa HPHT synthetic diamond substrate.

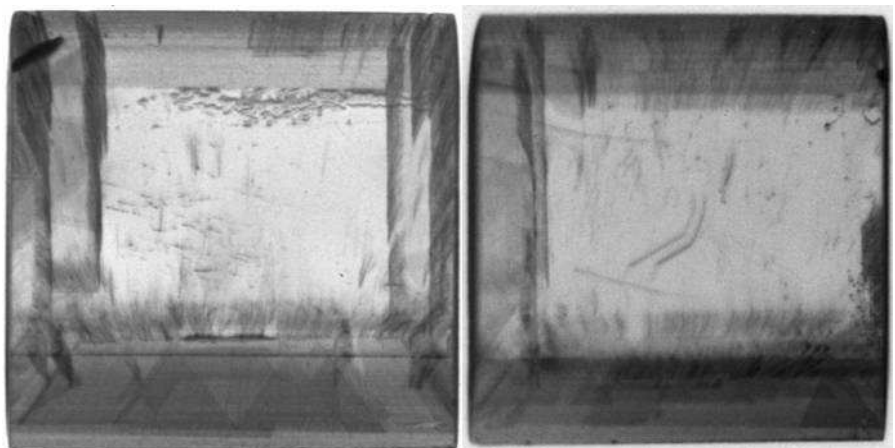

**Figure 6** Two {111} X-ray topographs of a CVD synthetic diamond layer on its HPHT synthetic diamond substrate showing that the central region is relatively free of extended defects.

The HPHT synthetic diamond substrate was laser sawn from the sample shown in figures 5 and 6. It was then re-polished and, after it had been fully characterized using X-ray topography, it was HPHT treated at $2400^{\circ}C$ for 15 minutes. It was then again re-polished to remove near-surface damage and re-characterized using X-ray topography in, as far as possible, an identical way. Figure 7 shows equivalent <111> topographs recorded before and after the treatment. There is clear evidence that much of the stacking fault contrast observed in a given topograph of the sample before treatment is absent from the equivalent topograph recorded after treatment. Comparison of the bottom left-hand corners of the second pair of topographs suggests that after HPHT treatment a dislocation has formed in a position corresponding to the edge of the removed stacking fault. It can also be seen that the shapes of dislocation lines have in some cases been changed by the treatment.

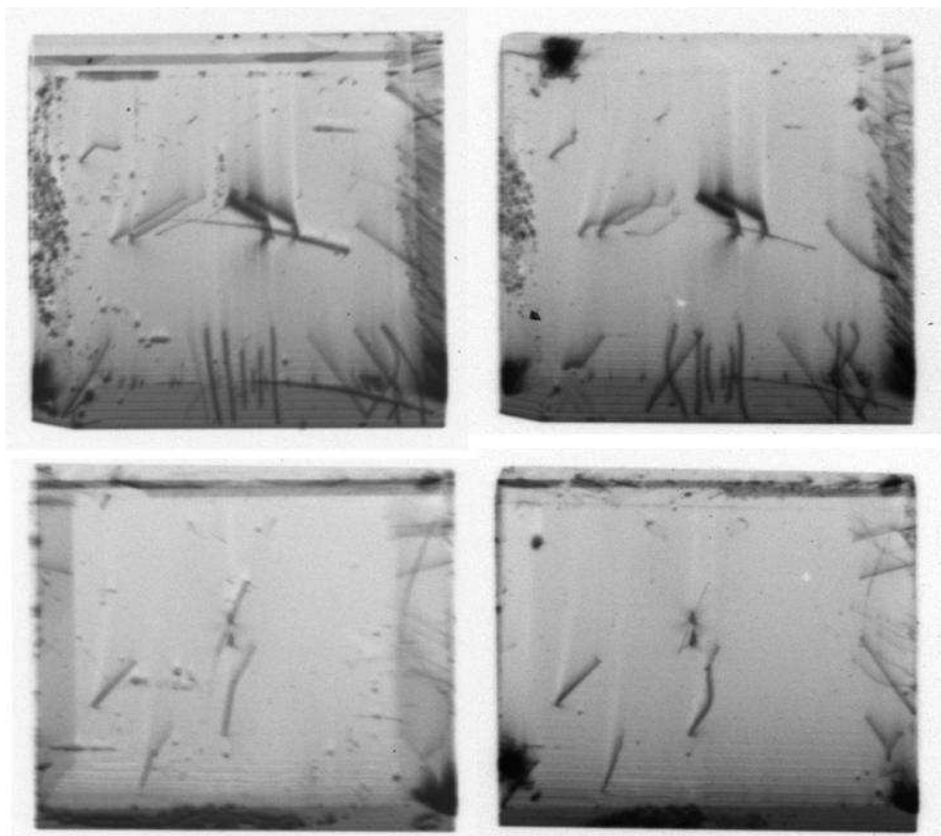

**Figure 7** The same two {422} X-ray projection topographs of a type IIa HPHT synthetic diamond sample recorded before (left) and after (right) HPHT treatment at 2400°C for 15 minutes.

Figure 8 shows four <111> projection topographs for a sample consisting of a layer of CVD diamond on its (001) type IIa HPHT synthetic substrate. It can be seen that, although there are some remaining substrate stacking faults that have acted as sources of dislocations, over much of the CVD layer the dislocation density is very low. In the central region, measuring approximately 2 mm x 2 mm, the dislocation density is less than 400 per $cm^{-2}$. Birefringence microscopy along the CVD growth direction indicated that $\Delta n$ was less than $10^{-5}$ in the central region of this sample but was significantly greater in the regions above dislocations that had nucleated as a result of stacking faults in the substrate.

## 4. Discussion

We have previously demonstrated that X-ray topography is a very useful technique for studying the extended defect content of single crystal CVD diamond samples. It should be clear from the results of this study that, by giving information about the extended defect content of HPHT synthetic diamond substrates on which CVD diamond is grown, X-ray topography can also help in the development of approaches to minimizing dislocation content of single crystal CVD diamond. It can help in the choice of the most suitable HPHT synthetic diamond samples, and also the most suitable regions of each for processing into substrates.

Type IIa HPHT synthetic substrates appear to offer a distinct advantage in allowing low dislocation density material to be grown above the (001) central sector. This sector is generally of high crystalline perfection, as both stacking faults and dislocations appear to be mainly confined to the {111} sectors. Clearly this helps towards the generation of high crystalline quality CVD material, but comparison with growth on type Ib HPHT synthetic substrates suggest that there is an additional advantage for CVD growth above {001} sectors of type IIa HPHT diamond. For growth on {001} type Ib substrates, significant numbers of mixed <001> dislocations were found to emanate from substrate polishing damage and groups of edge <100> dislocations nucleated at isolated points at the interface with the substrate. The number of these dislocations per unit area is seen to be significantly lower in this study and this is attributed to fact that a type IIa substrate has been used.

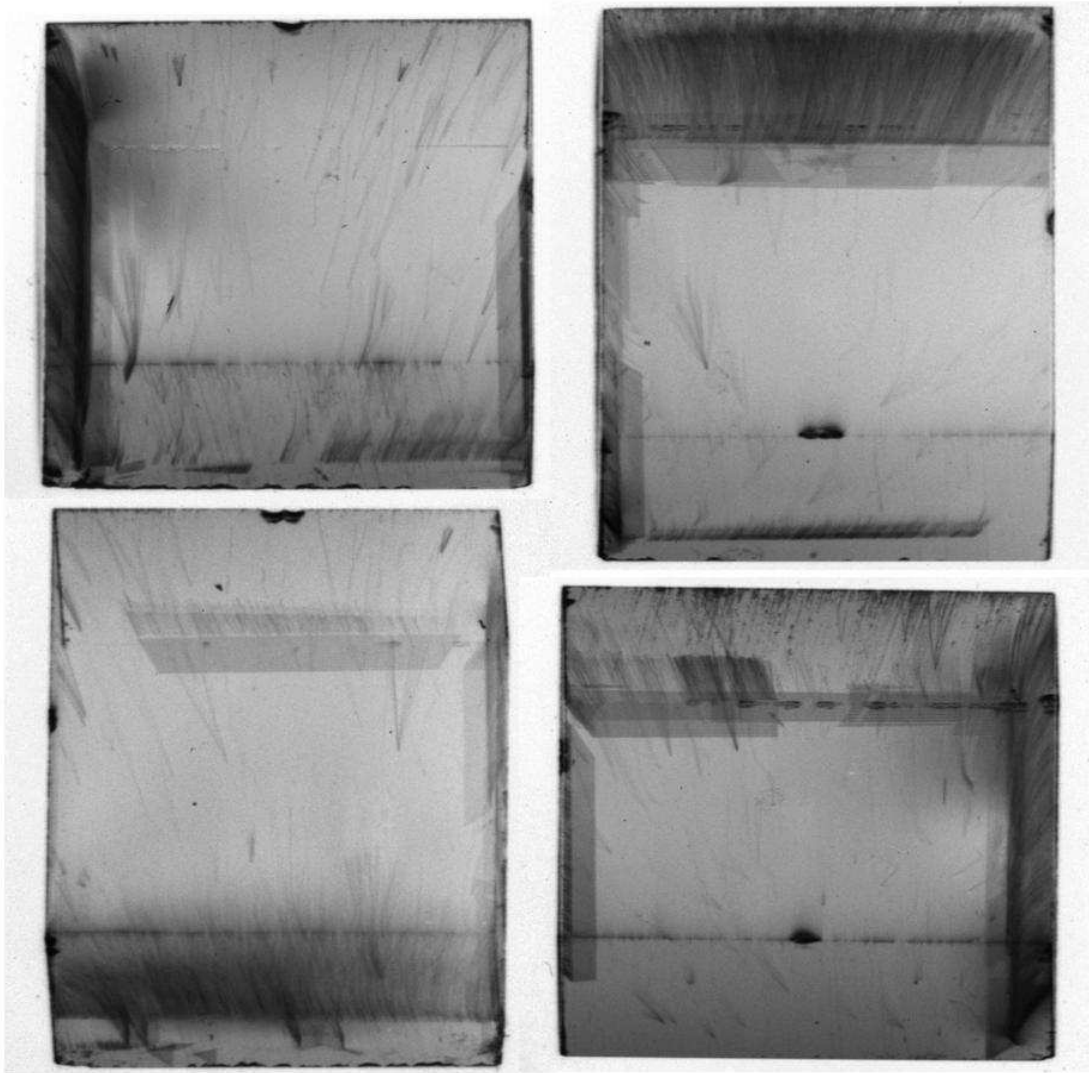

**Figure 8** Four {111} X-ray projection topographs of a single crystal CVD diamond layer on its type IIa HPHT synthetic diamond substrate. The central region shows a dislocation density that is less than 400 cm$^{-2}$.

There are various possible reasons for the apparent benefits from using type IIa substrates. Type Ib HPHT synthetic substrates tend to contain tens to hundreds of parts per million of nitrogen impurity and it is known that this increases the lattice parameter of the material significantly (Lang et al 1991). As a result, higher purity CVD material grown on such substrates is under tension at the interface and this may increase the tendency for nucleation of dislocations relative to what is the case when lattice parameters are better matched because of the use of type IIa substrates. Nitrogen impurity can change the mechanical properties of diamond and this in turn is likely to cause differences in the quality of type Ib and type IIa polished surfaces. This is relevant because it has been demonstrated that polishing damage at the surface on which CVD is initiated can act as a source of dislocations. In addition, type Ib and type IIa diamond are likely to etch at different rates in the CVD plasma environment and for a given process this could lead to different degrees of completeness in the removal of near-surface damage.

It is clear that substrate stacking faults do not tend to propagate into the CVD layer but instead they act as sources of dislocations. It has not been possible to determine the Burgers vectors of <001> dislocations that formed during the CVD process where stacking faults reached the substrate surface. This was because no clear extinct conditions could be identified. We know, however, that stacking faults tend to etch preferentially at the start of the CVD process, leaving grooves that are very likely to act as sources of many closely spaced dislocations, possibly with a range of different Burgers vectors, making extinction difficult to achieve. It is also possible that the dislocations formed may help to accommodate the stacking fault relative to the non-faulted material that grows above. The need to avoid substrate stacking faults suggests that, when producing (001) CVD substrates from HPHT synthetic diamond samples that have growth sector development similar to that illustrated in figure 1, it may be advantageous to use the region of the sample furthest from seed. Although this will not maximize the total area of the substrate, it will generally maximize the area that is free of stacking faults that can act as sources of dislocations when the CVD process is initiated.

This work has demonstrated that HPHT treatment can reduce stacking fault sources of dislocations in substrates used for homoepitaxial growth of single crystal CVD diamond. We propose that at a high enough temperature the partial dislocation bounding a stacking fault with fault vector $a/3<111>$ may split into a perfect dislocation and a Shockley partial with Burgers vectors of $a/2 <110>$ and $a/6 <112>$ respectively. This may be represented as: $a/3<111> = a/2<110> + a/6<112>$. Reduction in the area of such a stacking fault is energetically favourable because the reduction in energy of the stacking fault is greater than the increase in energy associated with the creation of the perfect dislocation. After this work had been carried out we became aware of earlier research (Antsygin et al 1998) in which stacking fault reduction in HPHT synthetic diamond by HPHT treatment had been demonstrated at similar temperatures to those that we have used.

We have also observed that HPHT treatment can cause dislocation lines to change shape. Many of the dislocations in the HPHT substrates appear to lie in {111} planes and often have <110> line direction. It seems likely that they move in {111} slip planes in response to deviations from non-hydrostatic stress in the HPHT treatment process.

## 5. Conclusions

When compared with previous X-ray topography studies, these results suggest that, for substrates for homoepitaxial CVD, carefully chosen high quality type IIa HPHT synthetic diamond offers significant advantages over type Ib material, primarily because of the low dislocation densities for the central (001) sector and the smaller number of dislocations that nucleate at the surface on which CVD is initiated. It is clear, however, that stacking faults in {111} sectors can act as an important source of dislocations and therefore care must be taken to minimize the extent to which stacking faults reach the initial growth surface. We have shown that HPHT treatment can be used to reduce the area of stacking faults in a way that may reduce this source of dislocations and therefore aid the production of low birefringence material of high crystalline perfection for both optical and electronic applications.

Although of importance for optical and electronic applications, the development of synthetic diamond material with low extended defect content should not be of great concern to gemmologists focused on identifying the origin of diamonds. Natural type IIa and type IIb diamonds, the closest equivalents in terms of impurity content to the CVD synthetic material of greatest interest for electronic applications, are severely plastically deformed as a result of conditions experienced in the earth's upper mantle or during eruption. For colourless type IIa natural diamond, this deformation, although giving rise to strain-related birefringence, has had little or no effect on the optical properties that help to make diamonds desirable in jewellery. Indeed, as this deformation bears witness to the impressive narrative that lies behind a natural diamond, it can stimulate the imagination in a way that for many may make the diamond more desirable. For the gemmologist, the important point is that diamonds deformed in this way have dislocation contents that are very different from CVD synthetic diamond, whether it has the standard dislocation content demonstrated in earlier work (Martineau et al 2004, Gaukroger et al 2008), with dislocation line directions predominantly aligned along the growth direction, or the low dislocation content demonstrated here.

Acknowledgements
The authors thank Chris Kelly and Matthew Sheehy for careful polishing of the substrates and samples studied in this work.